\documentstyle[12pt]{article}
\advance \topmargin by -\headheight
\advance \topmargin by -\headsep
     
\evensidemargin \oddsidemargin
\def\rf#1{(\ref{eq:#1})}
\def\lab#1{\label{eq:#1}}
\def\nonu{\nonumber}
\def\br{\begin{eqnarray}}
\def\er{\end{eqnarray}}
\def\be{\begin{equation}}
\def\ee{\end{equation}}

\def\lb{\lbrack}
\def\rb{\rbrack}

\def\({\left(}
\def\){\right)}
\def\lskip{\vskip\baselineskip\vskip-\parskip\noindent}

\def\bc{\begin{center}}
\def\ec{\end{center}}

\def\Tr{\mathop{\rm Tr}}                  % Tr - big trace

\newcommand\sbr[2]{\left\lbrack\,{#1}\, ,\,{#2}\,\right\rbrack} % commutator
\def\a{\alpha}
\def\b{\beta}

\def\d{\delta}

\def\l{\lambda}

\def\o{\over}

\def\p{\phi}
\def\P{\Phi}
\def\pa{\partial}

\def\pr{\prime}

\newcommand\twomat[4]{\left(\begin{array}{cc}  %%   2x2 matrix  %ESA
{#1} & {#2} \\ {#3} & {#4} \end{array} \right)}

\def\cA{{\cal A}}
\def\cB{{\cal B}}

\def\cD{{\cal D}}

\def\cG{{\cal G}}

\def\cL{{\cal L}}
\def\cM{{\cal M}}

\def\cP{{\cal P}}

\font \msb=msbm10 scaled \magstep1

\newcommand{\IC}{\mbox{\msb C} }

\newcommand{\IZ}{\mbox{\msb Z} }

\newfam\euffam
\font\sixeuf=eufm6
\font\eighteuf=eufm8
\font\twelveeuf=eufm10 scaled\magstep1
	\textfont\euffam=\twelveeuf
	\scriptfont\euffam=\eighteuf
  	\scriptscriptfont\euffam=\sixeuf

\def\proof{\par{\it Proof}. \ignorespaces} \def\endproof{{$\Box$}\par}

\def\vp{{\varphi}}

\newcommand{\ct}[1]{\cite{#1}}
\newcommand{\bi}[1]{\bibitem{#1}}
\newcommand\NPB[3]{{\sl Nucl. Phys.} {\bf B#1} (#2) #3}

\newcommand\CMP[3]{{\sl Commun. Math. Phys.} {\bf #1} (#2) #3}
\newcommand\PRD[3]{{\sl Phys. Rev.} {\bf D#1} (#2) #3}

\newcommand\PLB[3]{{\sl Phys. Lett.} {\bf #1B} (#2) #3}
\newcommand\JMP[3]{{\sl J. Math. Phys.} {\bf #1} (#2) #3}

\newcommand\JSM[3]{{\sl J. Soviet Math.} {\bf #1} (#2) #3}

\newcommand\JETPL[3]{{\sl  Sov. Phys. JETP Lett.} {\bf #1} (#2) #3}

\newcommand\JGP[3]{{\sl J. Geom. Phys.} {\bf #1} (#2) #3}

\def\kere{\mbox{\rm Ker (ad $E$)}}
\def\ime{\mbox{\rm Im (ad $E$)}}

\def\cgh{{\widehat {\cal G}}}

\begin{document}

\begin{titlepage}
\vspace{-1cm}
\noindent
\vskip .3in

\begin{center}

{\large\bf The complex sine-Gordon equation}
\end{center}
\begin{center}
{\large\bf as a symmetry flow of the AKNS hierarchy}
\end{center}
\normalsize
\vskip .4in

\begin{center}
{ H. Aratyn\footnotemark
\footnotetext{Work supported in part by NSF
(PHY-9820663)}}

\par \vskip .1in \noindent
Department of Physics \\
University of Illinois at Chicago\\
845 W. Taylor St.\\
Chicago, Illinois 60607-7059\\
\par \vskip .3in

\end{center}

\begin{center}
L.A. Ferreira{\footnotemark
\footnotetext{Work supported in part by CNPq}},
J.F. Gomes$^{\,2}$ and A.H. Zimerman$^{\,2}$

\par \vskip .1in \noindent
Instituto de F\'{\i}sica Te\'{o}rica-UNESP\\
Rua Pamplona 145\\
01405-900 S\~{a}o Paulo, Brazil
\par \vskip .3in

\end{center}

\begin{center}
{\large {\bf ABSTRACT}}\\
\end{center}
\par \vskip .3in \noindent

It is shown how the complex sine-Gordon equation arises as a symmetry flow
of the AKNS hierarchy. 
The AKNS hierarchy is extended by the ``negative'' symmetry flows forming the 
Borel loop algebra.
The complex sine-Gordon and the vector Nonlinear Schr\"{o}dinger
equations appear as lowest negative and second
positive flows within the extended hierarchy.
This is fully analogous to the well-known connection between 
the sine-Gordon and mKdV equations within the extended mKdV hierarchy.

A general formalism for a Toda-like symmetry occupying the ``negative''
sector of $sl(N)$ constrained KP hierarchy and giving rise to the negative
Borel $sl(N)$ loop algebra is indicated.

\end{titlepage}

A connection between the mKdV hierarchy and the sine-Gordon equation has been 
a recurrent theme in the soliton literature, see \ct{chodos,ds} and  
references therein.
As observed already in 1980 \ct{chodos}, the Hamiltonians of
the mKdV hierarchy remain conserved also with respect to the sine-Gordon flow.
This coincidence finds a natural explanation in the framework in which the mKdV hierarchy
is embedded in the extended hierarchy consisting of mutually commuting positive and negative
flows.
The positive part of the hierarchy comprises of the mKdV hierarchy while its negative 
counterpart contains the sine-Gordon equation and its own hierarchy of
differential equations.
The existence of two mutually compatible family of flows for every integrable
system is a reflection of the Riemann problem connected with two complementing solutions to the
underlying linear spectral problem. One solution method uses an expansion in negative powers of
the spectral parameter $\l$ and gives rise to the positive hierarchy while the other method
uses an expansion in the positive powers of $\l$ and gives rise to the
negative hierarchy.
In the present letter we show how to construct the hierarchy of the negative
flows and apply this method to the AKNS hierarchy.
The negative hierarchy is shown in the latter case to contain the complex
sine-Gordon equation, introduced in the context of the Lund-Regge model.

The approach we develop is a combination of the algebraic and pseudo-differential formalisms. 
In its general form it explains mutual commutativity of positive
and negative flows in the framework of constrained KP hierarchy which
contains AKNS model as a special case with the $sl(2)$ loop algebra and homogeneous
gradation \ct{AFGZ97}.

In the end of the letter we also comment on how our approach applies to the
$sl(n+1)$ mKdV type of hierarchies and we obtain the Toda type of hierarchies among 
the negative flows.

Let $\cgh = {\hat s}l (2)$ be a loop algebra with a graded
structure $ \cgh= \oplus_{n \in \IZ} \, \cgh_n$ given by a power series
expansion in the spectral parameter $\l$. This expansion defines an integral 
homogeneous gradation with respect to the gradation operator $d= \l d / d \l$.
The algebra $\cG = sl (2, \IC)$ has a standard basis 
$E_{\a} = \sigma_{+}$, $E_{-\a}= \sigma_{-}$ and $H = \sigma_{3}$.
We work within an algebraic approach to the integrable models based on
the linear spectral problem $L (\Psi)=0$ with a matrix Lax operator
$L= D_x + E + A$. Here, $E= \l \sigma_{3}/2$ is a semi-simple element 
of $\cgh$, chosen for simplicity to be a grade one element 
($E \equiv E^{(1)} \in \cgh_1$, $E^{(n)}= \l^n H/2 \in \cgh_n$). 
The matrix $A=q \sigma_{+}+r \sigma_{-}$ is the grade zero component 
of $\ime$ (see \ct{AFGZ97}).
Accordingly, the matrix operator $L$ for the AKNS hierarchy reads:
\be
L = \twomat{D+\l/2}{q}{r}{D-\l/2} = I \cdot D + {\l \o 2} H + 
q E_{\a} +r E_{-\a}
\lab{lopsl2}
\ee
here $D$ is the derivative with respect to $x$ acting to the right as an
operator according to the Leibniz rule.
In the corresponding formalism based on the pseudo-differential calculus
the equivalent spectral problem $\cL (\psi) = \l \psi$ is 
given in terms of the pseudo-differential Lax operator
$\cL = D -r D^{-1} q$.
The self-commuting isospectral flows ($n>0$): 
$ \pa_n r = B_n (r)$ and $\pa_n q = - B_n^* (q)$ with $B_n = 
(\cL^n)_{+}$ belong to the positive part of the AKNS hierarchy.
The conjugation ${}^*$ of $B_n$ is defined is such a way that $D=-D$
and $(AB)^*=B^*A^*$.
The second flow of the hierarchy:
\be
\pa_2 r = r_{xx}-2 q\,r^2 \;\;\; ; \;\;\; \pa_2 q = -q_{xx}+2 q^2\,r 
\lab{secflow}
\ee
gives the familiar vector non-linear Schr\"{o}dinger equation.

To define a ``negative part'' of the hierarchy
we need a matrix $M$ which arises  as a 
formal solution of the linear spectral problem:
\be
L (M) = \(\pa_x + E + A\) M = 0
\lab{transferm}
\ee
given in terms of the path ordered exponential:
\be
M = \cP e^{-\int^x \(  E + A\)dy}
\lab{treg}
\ee
where symbol $\cP$ denotes a path ordering.
Note, that all terms in the above exponential contain
only positive (and zero) grade generators.

The negative flows are induced by conjugation with the matrix $M$. 
To the element $X_{-n}= X \l^{-n}$ of  $\cgh_{-n}$ with $n>0$ we associate 
a flow:
\be 
\d^{(-n)}_X M \equiv  (M X_{-n} M^{-1})_{+} M  
\lab{sym-flows}
\ee
Direct calculation shows that these flows constitute a graded Borel loop algebra 
$\sbr{\d^{(-n)}_X}{\d^{(-m)}_Y}= \d^{(-n-m)}_{[X,Y]}$.
Their action on the grade-zero matrix $A$ is given by
\be
\d^{(-n)}_X A = - \lb (M X_{-n} M^{-1})_{-} \, , \,L \rb 
= - \lb (M X_{-n} M^{-1})_{-1} \, , \,E  \rb 
\lab{symm-a}
\ee
The flow generated by $X_{-1}= E^{(-1)}$ is of special interest and we now
provide for it a zero curvature formulation.
We choose the Gauss decomposition given by the following 
exponential of terms belonging to  zero grade subalgebra $\cgh_0 = sl (2)$:
\be
B = e^{ \chi E_{-\a}}\, e^{R H} \,e^{ \psi E_{\a}}
\lab{B-def}
\ee
%be such that $\pa_x B B^{-1} = A$. 
and define gauge potentials:
\be
\cA_{-} =  B E^{(-1)} B^{-1} %\lab{cam}\\
\;\; ;\;\; \cA_{+} =  -\pa_x B B^{-1}-E
\lab{cap}
\ee
In order to match the number of independent modes in the matrix $A$ we impose
two ``diagonal'' constraints $\Tr \(\pa_x B B^{-1} H \) = 0$ and
$\Tr \(B^{-1} {\bar \pa} B  H \)=0$ which effectively eliminate
$R$ in terms of $\psi$ and $\chi$.
In fact, those constraints reduce the zero grade subspace $\cgh_0 = sl (2)$
into the coset $sl(2)/U(1)$.
A more general and systematic construction for the affine non-abelian Toda
models in terms of the coset $sl(2) \otimes U(1)^{{\rm rank} \cG}/  U(1)$
is discussed in reference \ct{ime} where the models are constructed in terms of the
two-loop WZWN models \ct{2loop}.
Thus, after imposing these constraints $\cA_{+}$ becomes equal to 
$ -\pa_x B B^{-1} -E= -A-E = \pa_x M M^{-1}$
and the zero curvature condition: 
\be
\sbr{{\bar \pa}+\cA_{-}}{\pa_x +\cA_{+}} =
{\bar \pa} \cA_{+} - \pa_x \cA_{-} + \sbr{\cA_{-}}{\cA_{+}} =0
\lab{zcc}
\ee
holds for ${\bar \pa}= \d^{(-1)}_E$ as a consequence of \rf{symm-a}.

$B$ has been chosen so that after imposition of the constraints
$\pa_x B B^{-1} =  q E_{\a} +r E_{-\a}$.
Accordingly, we obtain the following representation for $q$ and $r$:
\be
q= {(\pa_x u) \o \Delta} e^R \quad; \quad 
r = (\pa_x {\bar u}) \, e^{-R}
\lab{qr-dict}
\ee
where
\be
u = \psi \, e^R \quad ; \quad {\bar u} = \chi \, e^R \quad ;
\quad \Delta = 1+ u\,{\bar u}
\lab{tipc-def}
\ee
with non-local field $R$ being determined in terms $u$ and ${\bar u}$ 
from the ``diagonal'' constraints :
\br
\Tr \(\pa_x B B^{-1} H \) &=& 0 \; \to \; \pa_x R = { {\bar u} \pa_x
u \o \Delta} \lab{par}\\
\Tr \(B^{-1} {\bar \pa} B  H \) &=& 0 \; \to \; {\bar \pa} R 
= { u {\bar \pa} {\bar u} \o \Delta} \lab{bpar}
\er
The zero curvature equations \rf{zcc}:
\br
{\bar \pa} q &=& {\bar \pa} \( {\pa_x u \o \Delta} e^R \)
= - 2 u e^R \lab{LRa}\\
{\bar \pa} r &=&  {\bar \pa} \(\pa_x {\bar u} \, e^{-R}\)=
- 2 {\bar u} \Delta e^{-R} \lab{LRb}
\er
together with eqs. \rf{par}-\rf{bpar} take now a form of the 
complex sine-Gordon equations \ct{lund,getman}:
% \br
% \pa_x \( { {\bar \pa} {\bar u} \o \Delta}\) + { {\bar u} \pa_x {\wti
% \psi} {\bar \pa} {\bar u} \o {\Delta}^2 } +2  {\bar u} &=& 0
% \lab{sGa}\\
% {\bar \pa}\( { \pa_x u \o \Delta}\) + { u \pa_x {\wti
% \psi} {\bar \pa} {\bar u} \o {\Delta}^2 } +2  u &=& 0
% \lab{sGb}
% \er
\br
 \pa_x   {\bar \pa} u + {u^* \pa_x u  {\bar \pa} u \o 1-u\, u^*}+ 2 u (1-u\,
u^*) &=& 0
 \lab{sGa}\\
 \pa_x   {\bar \pa} u^*  + {u\pa_x u^*   {\bar \pa} u^*  \o 1-u\, u^*}+ 2 u^* 
 (1-u\, u^*) &=& 0
 \lab{sGb}
\er
after substitution $u \to i u$ and ${\bar u} \to i u^* $.
Notice, that with the identification from \rf{qr-dict}-\rf{bpar} 
the $\cA_{+}$ component of gauge potentials are shared by
the AKNS and complex sine-Gordon theories.
Therefore, by gauge transforming $\cA_{+} $ in \rf{cap}
 into the $\kere$ we obtain simultaneous Hamiltonians
 for both complex sine-Gordon and AKNS models.

We now sketch a pseudo-differential approach to the study of ``negative''
flows developed in \ct{AGNP00}. Here we work with the AKNS Lax 
operator $\cL = D -r D^{-1} q$.
First, note that $\cL$ can be described as a ratio of two ordinary monic
differential operators as
$\cL = L_2 L_1^{-1}$, where $ L_1,L_2$ denote monic operators
$ L_1 = (D+ \vp_1^{\pr}+ \vp_2^{\pr})$ and 
$ L_2 = (D+ \vp_1^{\pr}) (D+ \vp_2^{\pr})$
of, respectively, order $1$ and $2$.
A monic differential operator $L_2$ is fully characterized
by elements of its kernel, $\p_1 = \exp (- \vp_2)$ and
$\p_2 = \exp (- \vp_2)\int^x \exp ( \vp_2-\vp_1)$.
Its inverse $L_2^{-1}$, is given by
$L_2^{-1} = \sum_{\a=1}^2 \p_{\a} D^{-1} \psi_{\a}$, where
$\psi_1 = - \exp ( \vp_1)\int^x \exp ( \vp_2-\vp_1)$
and $\psi_2 =  \exp ( \vp_1)$ are kernel elements of the conjugated
operator $L_2^{*}= (-D+ \vp_2^{\pr})(-D+ \vp_1^{\pr}) $, see \ct{UIC-97} and
references therein.
In this notation, $ \cL = D + L_2 (\exp ( -\vp_1-\vp_2))\,D^{-1} \exp (
\vp_1+\vp_2)$ and accordingly:
\be
q= \exp (\vp_1+\vp_2) \; \; ; \; \; r= \( \vp_1^{\pr \pr} - \vp_1^{\pr} 
\vp_2^{\pr} \) \exp (-\vp_1-\vp_2)
\lab{qr-phis}
\ee
Similarly, the inverse of $\cL$ is too given as a ratio of differential
operators
$\cL^{-1} = L_1  L_2^{-1}=  \sum_{\a=1}^2 L_1(\p_{\a}) D^{-1} \psi_{\a}$.
The functions $\P_{\a}^{(-1)} \equiv  L_1(\p_{\a})$ and
$ \Psi_{\a}^{(-1)} \equiv \psi_{\a}$ satisfy the same flow equations
as $r$ and $q$ with respect to the positive flows of the AKNS hierarchy.
We now extend the AKNS hierarchy by the ``negative'' flows
generated by the pseudo-differential operators \ct{AGNP00}:
\be
\cM^{(-n)}_{\cA} = \sum_{\a,\b=1}^2 \cA_{\a \b} \sum_{s=1}^n 
\P_{\b}^{(-n+s-1)} D^{-1} \Psi_{\a}^{(-s)}
\quad ;\quad n=1,2,3, {\ldots} 
\lab{cMn}
\ee
where $\P_{\a}^{(-n)}= \cL^{-n+1} (\P_{\a}^{(-1)})$ and
$\Psi_{\a}^{(-n)}= (\cL^*)^{-n+1} (\Psi_{\a}^{(-1)})$ are expressed entirely 
by the phase variables $\vp_1$ and $\vp_2$ of the AKNS hierarchy.
Furthermore, $\cA_{\a \b}$ is a constant $2\times 2$ matrix.
The corresponding ``negative'' symmetry flows are defined by:
\be
\cD^{(-n)}_{\cA} \cL = \sbr{\cM^{(-n)}_{\cA}}{\cL}
\lab{cd-def}
\ee
The following relations follow from \rf{cd-def} and determine flows on
$\P_{\a}^{(-m)},\Psi_{\a}^{(-m)}$:
\br
\cD^{(-n)}_{\cA} (\P_{\a}^{(-m)}) &=& \cM^{(-n)}_{\cA} \( \P_{\a}^{(-m)}\) - \sum_{\b=1}^2
\cA_{\a \b} \P_{\b}^{(-n-m)} \lab{cDPhi}\\
\cD^{(-n)}_{\cA} (\Psi_{\a}^{(-m)}) &=& - \(\cM^{(-n)}_{\cA}\)^{*} \(
\Psi_{\a}^{(-m)}\) +\sum_{\b=1}^2
\cA_{\b \a} \Psi_{\b}^{(-n-m)} \lab{cDPsi}
\er
These relations ensure that $\cD^{(-n)}_{\cA}$ span a graded Borel loop 
algebra: $ \sbr{\cD^{(-n)}_{\cA}}{\cD^{(-m)}_{\cB}}=
\cD^{(-n-m)}_{\sbr{\cA}{\cB}}$.
The flows $\cD^{(-n)}_{\cA}$ preserve the constrained structure of the
AKNS hierarchy and act on (adjoint) eigenfunctions $q$ and $r$
according to $\cD^{(-n)}_{\cA} (r) = \cM^{(-n)}_{\cA} (r)$
and $\cD^{(-n)}_{\cA} (q) = - \(\cM^{(-n)}_{\cA}\)^{*} \(q\)$, due to identities
$\cL  (\P_{\a}^{(-1)})=0 $ and $(\cL^*) (\Psi_{\a}^{(-1)})=0$.
It is interesting to note at this point that the generating functions $F_{\a} (\l)=
\sum_{n=1}^{\infty} \l^{n-1} \P_{\a}^{(-n)}$ and $G_{\a} (\l)=
\sum_{n=1}^{\infty} \l^{n-1} \Psi_{\a}^{(-n)}$ for $\P_{\a}^{(-n)}$ and
$\Psi_{\a}^{(-n)}$ are the solutions of the spectral
problems $\cL (F_{\a} (\l)) = \l F_{\a} (\l)$, $\cL^{*} (G_{\a} (\l)) = 
\l G_{\a} (\l)$.

We now present two of the main results of this paper.
First, the flows $\cD^{(-n)}_{\cA}$ commute with the isospectral
flows of the AKNS hierarchy.
This follows from \rf{cd-def} and the fact that $ \P_{\a}^{(-n)}$,
$\Psi_{\a}^{(-n)}$ are (adjoint) eigenfunctions with respect to isospectral
flows, i.e.  $\pa_n \P_{\a}^{(-m)} = B_n ( \P_{\a}^{(-n)})$ and
$ \pa_n \Psi_{\a}^{(-m)} = -B_n^* ( \Psi_{\a}^{(-m)} )$.
Accordingly, the flows $\cD^{(-n)}_{\cA}$ define the symmetry of the AKNS
hierarchy.
One can generalize this result to the case of the arbitrary constrained
KP model associated with the loop algebra ${\hat sl} (N)$ and with the Lax
operator $\cL= (\cL)_{+}+\sum_{i=1}^M \P_i D^{-1} \Psi_i$ with $M<N$ \ct{AFGZ97}.
It holds in that case that
the flows of the negative Borel loop algebra will commute with the
flows of the positive Borel loop algebra, which has recently been defined
for the constrained KP hierarchy in reference \ct{AGNP00}, which contains
several technical lemmas helpful for completing the proofs ommitted here. 
The final result is that \ct{AGNP00}
$\sbr{\cD^{(-n)}_{\cA}}{\cD^{(m)}_{\cB}}=0$ where ${\cD^{(m)}_{\cB}}$
are flows corresponding to the positive Borel loop algebra 
defined in \ct{AGNP00} with $\cB$ being a constant $M
\times M$ matrix, $n,m>0$, $\cA$ is a constant $N \times N$ matrix appearing
in a straightforward generalization of \rf{cMn} \ct{AGNP00}.

Secondly, the flows $\cD^{(-n)}_{\cA}$ defined in \rf{cd-def} for the
AKNS hierarchy coincide with the flows
$\d^{(-m)}_X$ defined by the matrix $M$ for $m=n >0$ and $X= \cA=\sigma_3$.
This observation provides an indirect proof for that the flows induced by
the conjugation with the matrix $M$ in \rf{sym-flows} and \rf{symm-a}
are the symmetry flows of the AKNS hierarchy and in particular commute 
with the isospectral flows.
We will ilustrate the identity $\cD^{(-n)}_{\cA} = \d^{(-n)}_{\cA}$ for
$\cA=\sigma_3$ and $n=1$.
{}From \rf{cDPhi}-\rf{cDPsi} we find
\be
\cD^{(-1)}_{\sigma_3} (\vp_1) = -2 \int^x e^{\vp_2-\vp_1}\, \int^x
\vp_1^{\pr} e^{\vp_1-\vp_2} \;\; ; \;\; 
\cD^{(-1)}_{\sigma_3} (\vp_2) = 2 \int^x e^{\vp_2-\vp_1}\, \int^x
\vp_2^{\pr} e^{\vp_1-\vp_2} 
\lab{trsigma3}
\ee
which, using expressions \rf{qr-phis}, leads to:
\br
\cD^{(-1)}_{\sigma_3} (q)&=& -2 e^{2\vp_1}\, \int^x
e^{\vp_2-\vp_1} \lab{trsq}\\
\cD^{(-1)}_{\sigma_3} (r)&=& -2\vp_1^{\pr}\, e^{-\vp_1-\vp_2}
\( 1+\vp_1^{\pr}e^{\vp_1-\vp_2}
 \int^x e^{\vp_2-\vp_1} \) \lab{trsr}
\er
Comparing with equations \rf{LRa}-\rf{LRb} we see that equality
${\bar \pa}=\cD^{(-1)}_{\sigma_3}$ holds provided we identify:
\be
R=\vp_1 \;\; ; \; \; u = e^{\vp_1}\,\int^x e^{\vp_2-\vp_1}  \;\; ; \; \;
{\bar u} = \vp_1^{\pr}\,e^{-\vp_2}
\lab{ident}
\ee
With representation \rf{ident} and transformations \rf{trsq}-\rf{trsr}
the constraints \rf{par}-\rf{bpar}
hold automatically and \rf{LRa}-\rf{LRb} are satisfied as well 
with ${\bar \pa}=\d^{(-1)}_{\sigma_3}=\cD^{(-1)}_{\sigma_3}$.
Similarly, we find that $\d^{(-1)}_{\sigma_{\pm}}=-\cD^{(-1)}_{\sigma_{\mp}}$
with:
\br
\cD^{(-1)}_{\sigma_{+}} (q) &=& - u^2 \;\; ; \;\; \cD^{(-1)}_{\sigma_{+}} (r) =
 - e^{-2 \vp_1} \Delta^2 \lab{cdsigp}\\
\cD^{(-1)}_{\sigma_{-}} (q) &=& e^{2 \vp_1} \;\; ; \;\; \cD^{(-1)}_{\sigma_{-}} (r) =
{\bar u}^2 \lab{cdsigm}
\er  
Due to the fact that we are dealing with a Borel loop algebra all the remaining 
symmetry flows can be found from the commutator relations involving known
lower grade flows.

We now comment on the special case of the generalized mKdV model associated with
the ${\hat sl}(n+1)$ algebra with the principal gradation \ct{ds}.
In the algebraic approach the Lax matrix $L=D+A+E$ contains
\be
E = E^{(1)} = \sum_{j=1}^{n}  E^{(0)}_{\a_{j}}
+  E^{(1)}_{-(\a_{1}+ \cdots+\a_{n})}\;\;\ ; \;\; 
A= \sum_{i=1}^n \(\vp^{\pr}_1+{\ldots} +\vp^{\pr}_i \) \a_i \cdot H
\lab{a2}
\ee
with $E$ and $A$ possessing grade $1$ and zero according to the principal
gradation defined by the charge $Q= (n+1)d+\sum_{i=1}^n \l_i \cdot H$, where
$\l_i$ are fundamental weights corresponding to the simple roots $\a_i$.
The solution to the linear problem $(D+A+E)(M)=0$ is given by the path
ordered exponentials \ct{bazhanov}:
\br
M &=& e^{\sum_{i=1}^n \(\vp_1+{\ldots} +\vp_i \) \a_i \cdot H}\,
\cP e^{\int^x \sum_{i=1}^n \( f_i E^{(0)}_{\a_{i}}
  +f_0  E^{(1)}_{-(\a_{1}+ \cdots+\a_{n})}\)} \lab{prinmono}\\
f_j &=& e^{-\sum_{j=1}^n K_{ji} \( \vp_1 +{\ldots} +\vp_i\)}
\;\;;\;\; f_0=e^{\sum_{i=1}^n (K_{1i}+{\ldots} +K_{ni}) \( \vp_1 +{\ldots} +\vp_i\)}
\nonu
\er
where $K_{ij}$ is a Cartan matrix of $sl(n+1)$.
Inserting $X= E^{(0)}_{-\a_{j}}$ with grade $-1$ into
\rf{symm-a} with $M$ from \rf{prinmono} we obtain
\be
\d^{(-1)}_{-\a_{j}} (\vp^{\pr}_j) = - e^{-\vp_j+\vp_{j+1}}
\;\; ;\;\; \d^{(-1)}_{-\a_{j}} (\vp^{\pr}_{j+1}) = e^{-\vp_j+\vp_{j+1}}
\;\; ;\;\; \d^{(-1)}_{-\a_{j}} (\vp^{\pr}_l)= 0 \;\; l \ne j,j+1
\lab{dajpj}
\ee
while for $X= E^{(-1)}_{(\a_{1}+ \cdots+\a_{n})}$
we obtain
\be
\d^{(-1)}_{(\a_{1}+ \cdots+\a_{n})} (\vp^{\pr}_1) = e^{2\vp_1+\vp_2+{\ldots}
+\vp_{n}}
\;\; ;\;\; \d^{(-1)}_{(\a_{1}+ \cdots+\a_{n})} (\vp^{\pr}_l)= 0 \;\; l>1
\lab{dppj}
\ee
These results give for the element $X= E^{(-1)}= \sum_{j=1}^{n}  E^{(0)}_{-\a_{j}}
+  E^{(-1)}_{(\a_{1}+ \cdots+\a_{n})}$
and ${\bar \pa}= \d^{(-1)}_{E}$ the affine Toda equation for $sl (n+1)$:
\be
\pa_x {\bar \pa} y_i = e^{\sum_{j=1}^n K_{ij} y_j}- e^{\sum_{j=1}^n K_{0j} y_j} \;\; ;\;\; 
y_i =  - \sum_{j=1}^i \vp_{j}
\lab{atoda}
\ee
with the extended Cartan matrix $K_{ab}$.
In the corresponding pseudo-differential approach the mKdV Lax operator 
$\cL= L_{n+1}=(D+\vp_1^{\pr})\cdots (D+\vp_{n+1}^{\pr})$ with the trace condition 
$\vp_1+{\ldots} +\vp_{n+1}=0$) is an ordinary differential operator. 
Let $\p_{\a} \in {\rm Ker} (L_{n+1})$,
$\psi_{\a} \in {\rm Ker} (L_{n+1}^*)$ with $\p_1= \exp(-\vp_{n+1})$
and $\psi_{n+1}= \exp (\vp_1)$. For $\cA_{\a\b}= \d_{\a,n+1} \d_{\b,1}$
the corresponding generator $\cM^{(-1)}_{\cA}= \p_1 D^{-1} \psi_{n+1}$
will induce according to relation \rf{cd-def} :
\be
\cD^{-1}_{\cA} (\vp_1^{\pr})= e^{\vp_1- \vp_{n+1}} \; \; ; \;\;
\cD^{-1}_{\cA} (\vp_{n+1}^{\pr})= -  e^{\vp_1- \vp_{n+1}}
\lab{toda}
\ee
with $\cD^{-1}_{\cA} (\vp_j^{\pr})= 0$ for $1<j<n+1$.
We recognize in \rf{toda} the Toda structure of eq. \rf{dppj}.
The other transformations of eqs. \rf{dajpj} follow by applying the
Darboux-B\"{a}cklund transformations $\vp_i \to \vp_{i+j}\, ,\, 1 \leq j
\leq n$ (modulo $n+1$).

{\it Outlook}.
We presented a concept of (non-local) Toda-like symmetries occupying the ``negative''
sector of the $sl(N)$ constrained KP hierarchy and giving rise to the negative
Borel $sl(N)$ loop algebra.
The case of $sl(2)$ (both homogeneous and principal gradations) has been
described in details for AKNS and mKdV hierarchies.
Details of the corresponding Toda like models of the $sl(3)$ constrained
KP hierarchies will be given elsewhere.
It is also of interest to establish similar negative flow structure for the
graded algebras connected with supersymmetric integrable models in order to
obtain  a new point of view on the supersymmetric Toda systems.
We also plan to describe relation of the negative Borel additional symmetry
loop algebra to the complete (centerless) Virasoro symmetry recently
established for the arbitrary constrained KP models \ct{AGNP00}.
It will also be of interest to establish a general tau-function realization valid
for both positive and negative sectors of the integrable models.

\lskip
{\bf Acknowledgements }
We are indebted to L. Dickey for useful comments on the manuscript.
HA thanks Fapesp for financial support and IFT-Unesp for hospitality.

%\lskip

\end{document}